\documentclass{article}
\usepackage{ijcai24}

\usepackage{times}
\usepackage{soul}
\usepackage{url}
\usepackage[hidelinks]{hyperref}
\usepackage[utf8]{inputenc}
\usepackage[small]{caption}
\usepackage{graphicx}
\usepackage{amsmath}
\usepackage{amsthm}
\usepackage{booktabs}
\usepackage{algorithm}
\usepackage{algorithmic}

\usepackage[switch]{lineno}

\usepackage{amssymb}
\usepackage{multirow}
\usepackage{booktabs}
\usepackage{amssymb}
\usepackage{color}
\usepackage[misc]{ifsym}

\title{MAEDiff: Masked Autoencoder-enhanced Diffusion Models for Unsupervised Anomaly Detection in Brain Images}

\author{
Rui Xu$^1$
\and
Yunke Wang$^1$\and
Bo Du$^{1}$\Letter\\
\affiliations
$^1$School of Computer Science, Wuhan University\\
\emails
\{rui.xu, yunke.wang, dubo\}@whu.edu.cn
}

\begin{document}

\maketitle

\begin{abstract}
    Unsupervised anomaly detection has gained significant attention in the field of medical imaging due to its capability of relieving the costly pixel-level annotation. To achieve this, modern approaches usually utilize generative models to produce healthy references of the diseased images and then identify the abnormalities by comparing the healthy references and the original diseased images. Recently, diffusion models have exhibited promising potential for unsupervised anomaly detection in medical images for their good mode coverage and high sample quality. 
    However, the intrinsic characteristics of the medical images, e.g. the low contrast, and the intricate anatomical structure of the human body make the reconstruction challenging. Besides, the global information of medical images often remain underutilized. To address these two issues, we propose a novel Masked Autoencoder-enhanced Diffusion Model (MAEDiff) for unsupervised anomaly detection in brain images. The MAEDiff involves a hierarchical patch partition. It generates healthy images by overlapping upper-level patches and implements a mechanism based on the masked autoencoders operating on the sub-level patches to enhance the condition on the unnoised regions. Extensive experiments on data of tumors and multiple sclerosis lesions demonstrate the effectiveness of our method.
\end{abstract}

\section{Introduction}

Segmenting lesions in medical images holds significant clinical importance. Yet the supervised deep learning techniques for this task often require extensive pixel-level annotations, limiting their applicability in tasks like anomaly detection. Therefore, unsupervised anomaly detection approaches are proposed as an alternative. Specifically, approaches using generative models have been explored and exhibit promising performance \cite{midl22/dae,ijcars21/3d_erasing,midl22/svae,mia22/c_vae,mia22/vq_vae,mp21/wgan,isbi21/inpainting}. Typically, the generative models are trained on healthy data to learn their underlying distributions. Then when presented with anomalous data, these models can detect deviations from the learned normal distribution. This is achieved by performing a pixel-by-pixel comparison between the reconstructed images, which are also called the healthy references, and their original counterparts.

Diffusion models \cite{ho2020denoising} have recently emerged as a powerful tool for unsupervised anomaly detection in brain images. Diffusion models work by incrementally adding noise to map images into latent spaces of identical dimensions. Then a model is trained to reverse the process for denoising, thereby restoring the original image from its noise-perturbed state. Compared to generative adversarial networks (GANs) \cite{NeurIPS14/gan}, diffusion models demonstrate superior mode coverage, avoiding the issue of mode collapse often seen in GANs where only a limited portion of the data distribution is captured \cite{mia21/ae_vae}. In contrast to autoencoders (AEs) and variational autoencoders (VAEs) \cite{iclr14/vae}, diffusion models produce samples of higher fidelity and greater detail, as AEs and VAEs tend to generate blurrier images due to the loss of spatial information \cite{mia21/ae_vae}. Approaches proposed in \cite{miccai22/classifier_guide,cvprw22/anoddpm,miccai22/cpdm,miccai22/latent_diff,arxiv23/ddpm_ddim} utilize diffusion models to reconstruct the entire brain images. In \cite{midl23/pddpm}, a patch-based denoising diffusion probabilistic model (DDPM) approach is presented to reconstruct the brain images by sliding patches.

However, some of these approaches struggle to generate the fine anatomical structures of the brain images due to their focus on reconstructing the entire images. Some works do not make full use of the contextual information, as the diffusion networks are not inherently designed for this purpose. Hence, in this work, we propose a novel unsupervised anomaly detection method named Masked Autoencoder-enhanced Diffusion Model (MAEDiff). In our MAEDiff, we divide the entire brain image into hierarchical patches. During training, the forward diffusion process is applied on one of the upper-level patches, while the backward diffusion process processes the entire image. Thereinto, we add a mechanism based on the Masked Autoencoder (MAE), which operates on the sub-level patches, to the diffusion U-Net \cite{NeurIPS21/guided_diffusion} for better condition on the unnoised regions. The training loss is computed within the noised patch. During inference, the forward and backward diffusion processes are applied sequentially to the overlapping upper-level patches, where the overlapping regions are averaged. 

We encapsulate our key contributions in the ensuing points:

\begin{itemize}
    \item We introduce a novel Masked Autoencoder-enhanced Diffusion Model (MAEDiff) for unsupervised anomaly detection in brain images, which involves a hierarchical patch partition strategy.
    \item We propose to noise and denoise by overlapping upper-level patches. We integrate a mechanism based on Masked Autoencoder (MAE) into the diffusion U-Net, working on the sub-level patches.
\end{itemize}

Compared with the prior approaches, the results of extensive experiments on three public brain image datasets demonstrate the superiority of our proposed MAEDiff in anomaly detection and image reconstruction.

\section{Related Works}

\subsection{Anomaly Detection}

Generative model-based anomaly detection approaches have gained enormous traction in the research community for their dispense with costly pixel-level annotations. Among these approaches, autoencoders (AEs) and variational autoencoders (VAEs), generative adversarial networks (GANs), and the more recent diffusion models stand out as the most prominent. Given the tendency of autoencoders (AEs) to produce somewhat blurred reconstructions, there has been a focus among researchers on enhancing spatial detail retention. This enhancement is achieved through incorporating strategies like spatially enriched latent dimensions \cite{miccai18/enrich_latent}, employing multi-resolution techniques \cite{miccai20/multi_resolution}, integrating skip connections with dropout \cite{isbi20/skip_connection}, and leveraging the denoising task as a form of regularization \cite{midl22/dae}. In a similar vein, for variational autoencoders (VAEs), various methods have been proposed, including spatial erasure techniques \cite{midl19_/context_vae}, exploiting 3D data characteristics \cite{ijcars21/3d_erasing,midl22/svae}, and other innovative approaches \cite{mia20/norm_prior,mia22/c_vae,mia22/vq_vae}, to overcome their inherent limitations.
In contrast, to face the challenges of mode collapse and lack of anatomical fidelity in GANs \cite{mia19/f-anogan}, Wasserstein GANs \cite{mp21/wgan} and inpainting \cite{isbi21/inpainting} techniques have been employed as solutions. Recently, diffusion models have emerged as a potent alternative, effectively addressing both these issues simultaneously, thereby gaining prominence in the field of anomaly detection \cite{miccai22/classifier_guide,cvprw22/anoddpm,miccai22/cpdm,miccai22/latent_diff,arxiv23/ddpm_ddim}. Specifically, a novel approach employing a patch-based DDPM has been introduced in \cite{midl23/pddpm}, tailored to enhance the reconstruction accuracy of brain images.

\subsection{Denoising Diffusion Models}

The advent of denoising diffusion models marks a significant milestone in the generation of diverse and high-resolution conditional images, as evidenced by studies \cite{icml15/nonequil,NeurIPS21/guided_diffusion,ho2020denoising}. These models employ a forward process based on Gaussian diffusion, followed by a reverse generation mechanism. Through this approach, denoising diffusion models progressively enhance the clarity of an image initially derived from Gaussian noise. 
This technique has shown remarkable efficacy in generating rich, text-conditioned imagery \cite{arxiv22/h_text,icml22/glide,cvpr22/LDM} and even extends to video creation \cite{iclr23/phenaki,iclr23/makeavideo}.

\subsection{Masked Autoencoders}

The concept of masked autoencoders initially emerged through the development of stacked autoencoders \cite{jmlr10/stacked_ae} and the application in image inpainting tasks \cite{cvpr16/convinpainting} utilizing ConvNets. The advent of Vision Transformers (ViT) \cite{ircl21/vit} has reignited interest in masked prediction, drawing inspiration from the impactful strides made by BERT \cite{naacl19/bert} in the field of natural language processing (NLP). In NLP, BERT excels at masked language modeling, demonstrating scalability and adaptability to various tasks.

\begin{figure*}[!t]
  \centering
  \includegraphics[width=7in]{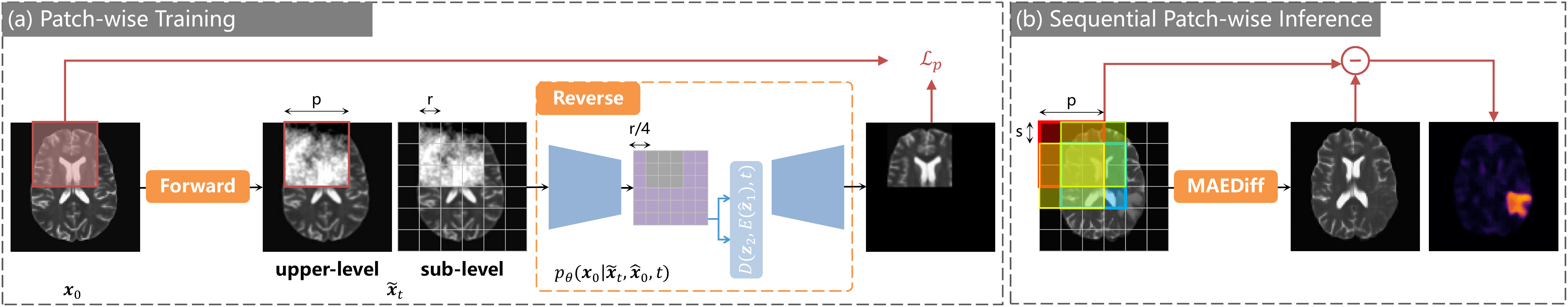}
  \caption{Overall pipeline of the proposed Masked Autoencoder-enhanced Diffusion Model (MAEDiff), where a hierarchical partition strategy is utilized. The input image $\boldsymbol{x}_{0}$ is first divided into larger upper-level $p \times p$ patches, and then further into smaller sub-level $r \times r$ grids. (a) In the training phase, one patch is randomly selected for patch-wise reconstruction. Specifically, the selected patch is diffused by the forward process, and reconstructed by the reverse process using the partially perturbed $\tilde{\boldsymbol{x}}_{t}$. The sub-level division mainly works on the feature map to enhance the condition on the visible (unnoised) region $\hat{\boldsymbol{x}}_{0}$. (b) In the testing phase, the patch-wise reconstruction is performed across the image by sliding horizontally and vertically at a step size of $s$. The anomaly score map is obtained by comparing the original diseased image with the reconstructed healthy reference pixel by pixel.
}
  \label{fig:patchify}
\end{figure*}

\section{Methodology}
In this section, we first briefly review the typical denoising diffusion probabilistic model (DDPM) \cite{ho2020denoising} and then introduce our Masked Autoencoder-enhanced Diffusion Model (MAEDiff) for unsupervised anomaly detection in brain images. 
MAEDiff hierarchically partitions the input image: initially dividing it into upper-level larger patches (can be overlapping), and then further subdividing these patches into smaller sub-level patches (not overlapping). For clarity, we denote the \textbf{upper-level} patches as \textit{patches} and the \textbf{sub-level} patches as \textit{grids}. For the diffusion process, we sample patches for individual reconstruction by moving over the patches (Sec.~\ref{sec:patch-based_scheme}). Regarding the generator architecture, we incorporate a masked autoencoder (MAE) \cite{cvpr22/mae} mechanism to operate on the grids. This integration into the diffusion U-Net \cite{NeurIPS21/guided_diffusion} is able to achieve better condition on the visible regions (Sec.~\ref{sec:mae_unet}).

\subsection{Preliminaries}
The DDPM~\cite{ho2020denoising} is a generative model specified by a $T$-step Markov chain. In DDPM, the forward diffusion process involves corrupting data through introducing Gaussian noise at each step. Following this paradigm, the diffusion model then learns the reverse denoising process, which is used to retore the clean image from a pure noise.

\paragraph{Forward Diffusion.} 
Given $\boldsymbol{x}_0\sim p(\boldsymbol{x}_0)$, the forward diffusion process 
progressively adds Gaussian noise to corrupt the data at each step according to $q(\boldsymbol{x}_t|\boldsymbol{x}_{t-1})$,
\begin{align}
    q(\boldsymbol{x}_t|\boldsymbol{x}_{t-1})=\mathcal{N}(\boldsymbol{x}_t;\sqrt{1-\beta_t}\boldsymbol{x}_{t-1},\beta_t\boldsymbol{I}),
    \label{eq:forward_diffusion_1}
\end{align}
where $\beta_t\in(0,1)$ denotes the pre-defined variance that is monotonically increasing with respect to timestep $t$. Given Eq. (\ref{eq:forward_diffusion_1}), the data distribution $q(\boldsymbol{x}_t|\boldsymbol{x}_0)$ at arbitrary timestep $t$ can be defined as,
\begin{equation}
    q(\boldsymbol{x}_t|\boldsymbol{x}_0) = \mathcal{N}(\boldsymbol{x}_t;\sqrt{\bar{\alpha}_t}\boldsymbol{x}_0,(1-\bar{\alpha}_t)\boldsymbol{I}),
    \label{eq:forward_diffusion_2}
\end{equation}
which can be also expressed in its closed form,
\begin{equation}
    \boldsymbol{x}_t = \sqrt{\bar{\alpha}_t}\boldsymbol{x}_0+\sqrt{1-\bar{\alpha}_t}\boldsymbol{\epsilon}_t
    \label{eq:forward_diffusion_3},
\end{equation}
where $\bar{\alpha}_t=\prod_s^t (1-\beta_s)$ and $\boldsymbol{\epsilon}_t\sim\mathcal{N}(0,1)$.
For $t=T$, the forward diffusion process will transform $\boldsymbol{x}_0$ to a pure Gaussian noise as $\sqrt{\bar{\alpha}_t}\rightarrow0$.

\paragraph{Reverse Diffusion.} 
Based on the forward diffusion process, the reverse process starts from an isotropic Gaussian noise $\boldsymbol{x}_T$ and gradually remove noises via $T$ steps to obtain a target data as follows,
\begin{align}
    \boldsymbol{x}_0 &\sim p_\theta(\boldsymbol{x}_T)\prod^T_{t=1}p_\theta(\boldsymbol{x}_{t-1}|\boldsymbol{x}_t), \\ p_\theta(\boldsymbol{x}_{t-1}&|\boldsymbol{x}_t)=\mathcal{N}(\boldsymbol{x}_{t-1};\boldsymbol{\mu}_\theta(\boldsymbol{x}_t,t),\boldsymbol{\Sigma}_\theta(\boldsymbol{x}_t,t)),
\end{align}
where $p_\theta(\boldsymbol{x}_{t-1}|\boldsymbol{x}_t)$ is modeled as a Gaussian. By applying the reformulation and Bayes rules, the ground-truth denoising transition is tractable via reverse conditional probability $q(\boldsymbol{x}_{t-1}|\boldsymbol{x}_t,\boldsymbol{x}_0) = \mathcal{N}(\boldsymbol{x}_{t-1};\boldsymbol{\mu}_q(\boldsymbol{x}_t,\boldsymbol{x}_0),\boldsymbol{\Sigma}_q(\boldsymbol{x}_t,\boldsymbol{x}_0))$ when conditioned on $\boldsymbol{x}_0$. Through further reparametering, $\boldsymbol{\mu}_q$ can be connected to the noise vector $\boldsymbol{\epsilon}_0$ and we can parameterize $\boldsymbol{\mu}_q$ as,
\begin{equation}
    \boldsymbol{\mu}_q(\boldsymbol{x}_t,\boldsymbol{\epsilon}_0)=\frac{1}{\sqrt{\alpha_t}}\big(
    \boldsymbol{x}_t - \frac{\beta_t}{\sqrt{1-\bar{\alpha}_t}}\boldsymbol{\epsilon}_0)
    \big).
    \label{eq:reverse_diffusion_4}
\end{equation}
The variance $\boldsymbol{\Sigma}_q(\boldsymbol{x}_t,t)$ is normally fixed and set to be $\frac{1-\bar{\alpha}_{t-1}}{1-\bar{\alpha}_t}\beta_t\boldsymbol{I}$. By matching $\boldsymbol{\mu}_\theta$ closely to $\boldsymbol{\mu}_q$, 
the model $\boldsymbol{\epsilon}_\theta$ aims to predict the noise vector $\boldsymbol{\epsilon}_0$, and thus can be optimizing by the objective $\mathcal{L}=\mathbb{E}[||\boldsymbol{\epsilon}_0-\boldsymbol{\epsilon}_\theta(\boldsymbol{x}_t,t)||^2]$.
This mean squared error objective is simplified from the variational bound. Except for predicting the noise, the model can also be designed to predict the denoised data $\boldsymbol{x}_0$ directly~\cite{arxiv22/h_text} and the objective can be written as,
\begin{align}
    \mathcal{L}=\mathbb{E}\big[||\boldsymbol{x}_0 - \boldsymbol{x}^{rec}_0||^2\big], \ s.t. \ \boldsymbol{x}^{rec}_0=p_\theta(\boldsymbol{x}_t, t)
    \label{eq:df_optim}
\end{align}
In this work, we replace $l2$ loss in Eq. (\ref{eq:df_optim}) with $l1$ loss to train the predictor $p_\theta$. During the inference, typical sampling of DDPM needs to go through all $T$ steps to reconstruct the clean data. As this way of sampling leads to high computational cost, we directly estimate $\boldsymbol{x}_0^{rec}$ via a trained $p_\theta(\boldsymbol{x}_t,t)$ at a fixed timestep $t$.
\paragraph{Conditional DDPM.} 
When it comes to image-to-image translation, there are often image pairs available, \textit{e.g.}, $(\boldsymbol{x}_0,\hat{\boldsymbol{x}}_{0})$ (In this work, the $\boldsymbol{x}_0$ is the entire input image, whereas the $\hat{\boldsymbol{x}}_{0}$ is the visible region). Under such a case, while the forward diffusion process remains unchanged, the reverse process of the diffusion model should additionally be conditioned on $\hat{\boldsymbol{x}}_{0}$. Therefore, the noise estimation network becomes $\boldsymbol{\epsilon}_{\theta}(\boldsymbol{x}_t,\hat{\boldsymbol{x}}_{0},t)$ and the image estimation network is $p_{\theta}(\boldsymbol{x}_0 \vert \boldsymbol{x}_t,\hat{\boldsymbol{x}}_{0},t)$.

\subsection{Patch-based Diffusion Scheme}
\label{sec:patch-based_scheme}
Brain images typically contain a substantial amount of high-frequency details, including the tissue boundaries and fine anatomical structures. Most existing generative approaches estimate the whole brain image and neglect this characteristic. Recently, patch-based DDPMs \cite{cvpr22/repaint,pami23/pddpm,midl23/pddpm} have been proposed for several generative tasks. These patch-based approaches are particularly beneficial for extracting local features, enabling a more effective capture of high-frequency details and facilitating local adjustments. However, they do not fully leverage the contextual information beyond the reconstructed region. Differently, we propose a novel hierarchical patch partition method and perform the forward and backward diffusion process in a patch-wise manner.

As illustrated in Figure~\ref{fig:patchify}, we divide an image $\boldsymbol{x}_{0} \in \mathbb{R}^{C \times H \times W}$ into non-overlapping $r \times r$ \textbf{sub-level} \textit{grids} ($r \textless p$), from which potentially overlapping $p \times p$ \textbf{upper-level} \textit{patches} can be sampled with uniform spacing. The spacing can be set to $s$ units ($s$ is divisible by $r$) along both the horizontal and vertical axes. For example, the partition in Figure~\ref{fig:patchify} results in $N = \frac{HW}{r^{2}}$ grids and $K = \frac{(H-p+s)(W-p+s)}{s^{2}}$ patches.

The overall pipeline of our method is depicted in Figure~\ref{fig:patchify}. The training procedure is outlined in Algorithm~\ref{alg:training}. During the training, we randomly select one $p \times p$ patch to diffuse, referred to as the noised patch, and the area outside the selected patch is called the visible region. 
Specifically, we define a binary mask $\boldsymbol{M}_{\boldsymbol{p}}$ with the same dimensions as the input image $\boldsymbol{x}_{0}$. In this mask, pixels inside the selected patch position are assigned a value of 1, while all the other pixels are set to 0. The partly noised image can be obtained by

\begin{equation}
    \tilde{\boldsymbol{x}}_{t} = \boldsymbol{x}_{t} \odot \boldsymbol{M}_{\boldsymbol{p}} + \boldsymbol{x}_{0} \odot \neg \boldsymbol{M}_{\boldsymbol{p}},
\end{equation}

where $\odot$ represents element-wise multiplication. Recall that we want to reconstruct the image in a patch-wise fashion and fully leverage the contextual information, so the entire $\tilde{\boldsymbol{x}}_{t}$ is input to our generator, and the visible region $\hat{\boldsymbol{x}}_{0}$ also serves as a conditioning input. In other words, the denoised image $\boldsymbol{x}_{0}^{rec}$ is generated by sampling from $p_{\theta}(\boldsymbol{x}_{0} \vert \tilde{\boldsymbol{x}}_{t},\hat{\boldsymbol{x}}_{0},t)$. Our training objective, specifically focused on the noised region, is defined as

\begin{equation}
    \mathcal{L}_{p} = \vert \boldsymbol{x}_{0} - \boldsymbol{x}_{0}^{rec} \vert \odot \boldsymbol{M}_{\boldsymbol{p}}.
\end{equation}

\begin{algorithm}[tb]
    \caption{Patch-wise Training}
    \label{alg:training}
    \begin{algorithmic}[1] 
        \REPEAT
            \STATE $\boldsymbol{x}_{0} \sim p(\boldsymbol{x}_{0})$
            \STATE $t \sim {\rm Uniform}(\{1,2,\cdots,T-1,T\})$
            \STATE Randomly generate simplex seed
            \STATE $\boldsymbol{\epsilon} \sim {\rm Simplex}(\nu = 2^{-6},N = 6,\gamma = 0.8)$
            \STATE $\boldsymbol{x}_t = \sqrt{\bar{\alpha}_t}\boldsymbol{x}_0+\sqrt{1-\bar{\alpha}_t}\boldsymbol{\epsilon}$
            \STATE Ramdomly sample a binary mask $\boldsymbol{M}_{\boldsymbol{p}}$
            \STATE $\tilde{\boldsymbol{x}}_{t} = \boldsymbol{x}_{t} \odot \boldsymbol{M}_{\boldsymbol{p}} + \boldsymbol{x}_{0} \odot \neg \boldsymbol{M}_{\boldsymbol{p}}$ \textcolor{black}{\COMMENT{patch-wise diffusion}}
            \STATE $\hat{\boldsymbol{x}}_{0} = \boldsymbol{x}_{0} \odot \neg \boldsymbol{M}_{\boldsymbol{p}}$ \textcolor{black}{\COMMENT{visible region}}
            \STATE Take gradient descent step on \\
            $\nabla_{\theta} \vert \boldsymbol{x}_{0} - p_{\theta}(\tilde{\boldsymbol{x}}_{t},\hat{\boldsymbol{x}}_{0},t) \vert \odot \boldsymbol{M}_{\boldsymbol{p}}$
            \UNTIL converged
        \RETURN {$\theta$}
    \end{algorithmic}
\end{algorithm}

\begin{algorithm}[tb]
    \caption{Sequential Patch-wise Inference}
    \label{alg:inference}
    \begin{algorithmic}[1] 
    
            \STATE $\boldsymbol{x}_{0} \sim A(\boldsymbol{x}_{0})$
            \STATE Randomly generate simplex seed
            \STATE $\boldsymbol{\epsilon} \sim {\rm Simplex}(\nu = 2^{-6},N = 6,\gamma = 0.8)$
            \STATE $\boldsymbol{x}_t = \sqrt{\bar{\alpha}_t}\boldsymbol{x}_0+\sqrt{1-\bar{\alpha}_t}\boldsymbol{\epsilon}$
            \STATE $\boldsymbol{x}_{0}^{rec}=\boldsymbol{0}$ and $\boldsymbol{M}=\boldsymbol{0}$
            \FOR{$k=1,\cdots,K$}
                \STATE Sample a binary mask $\boldsymbol{M}_{\boldsymbol{p}_{k}}$    \STATE $\tilde{\boldsymbol{x}}_{t} = \boldsymbol{x}_{t} \odot \boldsymbol{M}_{\boldsymbol{p}_{k}} + \boldsymbol{x}_{0} \odot \neg \boldsymbol{M}_{\boldsymbol{p}_{k}}$ 
                \STATE $\hat{\boldsymbol{x}}_{0} = \boldsymbol{x}_{0} \odot \neg \boldsymbol{M}_{\boldsymbol{p}_{k}}$
                \STATE $\boldsymbol{x}_{0}^{rec}=\boldsymbol{x}_{0}^{rec} + p_{\theta}(\tilde{\boldsymbol{x}}_{t},\hat{\boldsymbol{x}}_{0},t) \odot \boldsymbol{M}_{\boldsymbol{p}_{k}}$
                \STATE $\boldsymbol{M}=\boldsymbol{M}+\boldsymbol{M}_{\boldsymbol{p}_{k}}$
            \ENDFOR
            \STATE $\boldsymbol{x}_{0}^{rec} = \boldsymbol{x}_{0}^{rec} \oslash \boldsymbol{M}$ \textcolor{black}{\COMMENT{average the overlapping region}}
            \STATE $\Delta_{AS} = \vert \boldsymbol{x}_{0} - \boldsymbol{x}_{0}^{rec} \vert$ \textcolor{black}{\COMMENT{pixel-wise anomaly score}}
        \RETURN {$\Delta_{AS}$}
    \end{algorithmic}
\end{algorithm}

Algorithm~\ref{alg:inference} delineates the inference steps. During inference, the $K$ patches are sampled in sequence for the patch-wise reconstruction. Each selected patch is partially diffused to a predetermined timestep, and then reconstructed. Upon the reconstruction of all $K$ patches, these reconstructed segments are meticulously assembled based on their respective positions in the input image, thereby restoring the original dimensionality of $\boldsymbol{x}_{0}$. To mitigate merging artifacts, the overlapping regions between patches are averaged.

In order to incorporate the MAE \cite{cvpr22/mae} into the diffusion U-Net \cite{NeurIPS21/guided_diffusion}, we apply the non-overlapping $\frac{r}{4} \times \frac{r}{4}$ grid division to the feature map $\boldsymbol{f} \in \mathbb{R}^{C \times \frac{H}{4} \times \frac{W}{4}}$ extracted by the U-Net encoder, which also results in $N = \frac{HW}{r^{2}}$ grid cells. Thus, the grid cells corresponding to the selected patch are considered masked, while the remaining grid cells are considered visible $\hat{\boldsymbol{f}}$.

\begin{figure}[!t]
  \centering
  \includegraphics[width=3.5in]{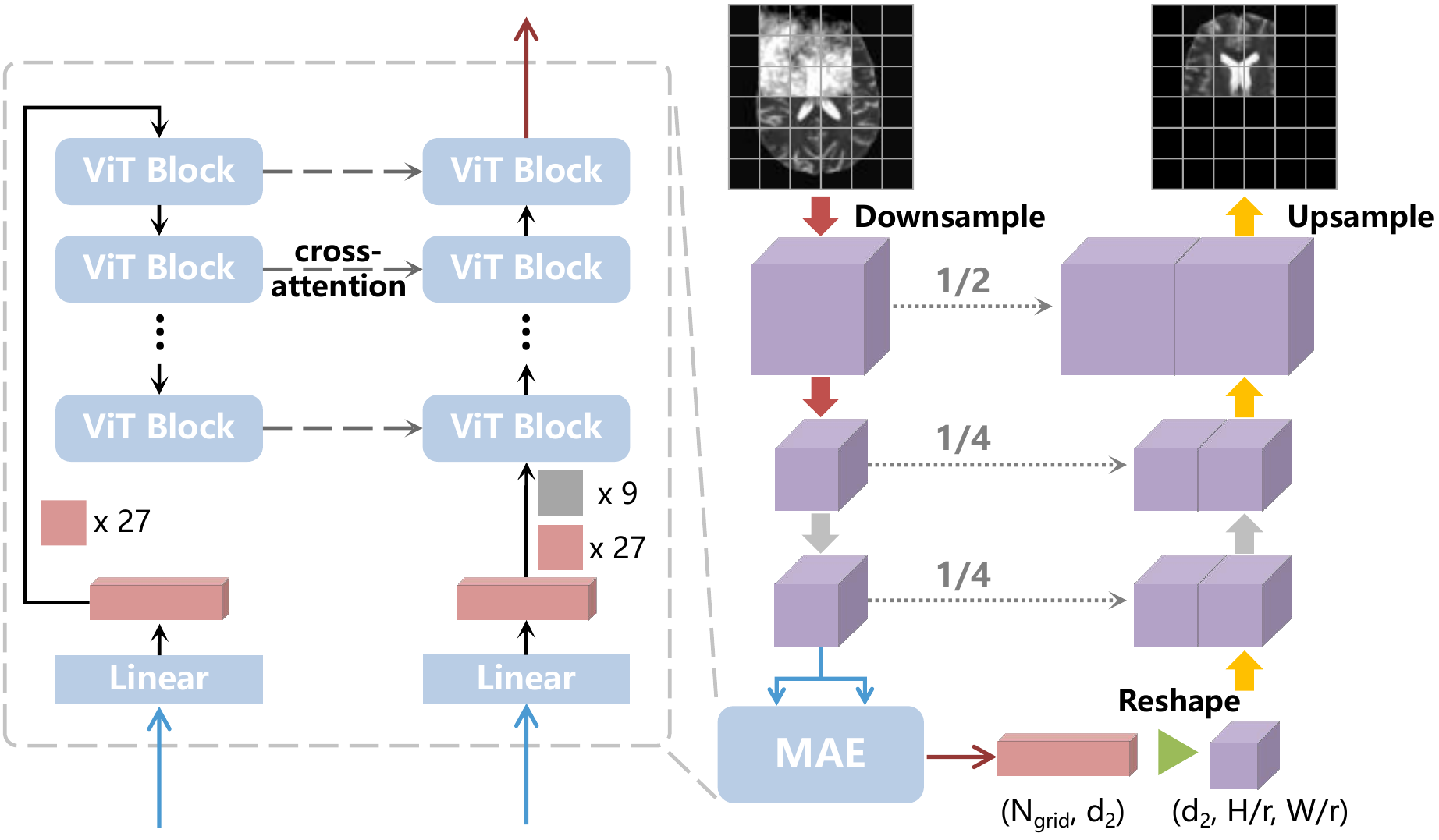}
  \caption{The generator architecture of our MAEDiff, which is a typical diffusion U-Net integrated with an MAE-like mechanism. Thereinto, the feature map extracted by the U-Net encoder is fed into the MAE module. The visible feature map is processed by the MAE encoder, and then processed along with the entire feature map by the MAE decoder. The output of the MAE module is reshaped and upsampled to be compatible with the remaining U-Net.
}
  \label{fig:maediff_unet}
\end{figure}

\subsection{Masked Autoencoder-enhanced U-Net}
\label{sec:mae_unet}

In diffusion models \cite{NeurIPS21/guided_diffusion}, including the patch-based diffusion models \cite{cvpr22/repaint,pami23/pddpm,midl23/pddpm}, the U-Net \cite{miccai15/unet} backbones operate directly on the entire input image. However, this is sub-optimal for the patch-based DDPMs, as the original U-Net architecture is not explicitly designed to handle the interaction between the noised region and the visible region. Thus to better reconstruct the noised patch from the visible region, we integrate an MAE \cite{cvpr22/mae} mechanism into the U-Net, as illustrated in Figure~\ref{fig:maediff_unet}.

Different from \cite{cvpr22/mae}, we choose to operate the MAE on the feature map $\boldsymbol{f} \in \mathbb{R}^{C \times \frac{H}{4} \times \frac{W}{4}}$, which is extracted by the U-Net encoder. As previously mentioned, we apply a non-overlapping 
$\frac{r}{4} \times \frac{r}{4}$ grid division to the feature map, resulting in $N = \frac{HW}{r^{2}}$ grid cells. Grid cells that correspond to the noised patch are treated as masked, while the others are considered visible $\hat{\boldsymbol{f}}$.

\paragraph{MAE Encoder.} Following \cite{cvpr22/mae}, the encoder comprises Vision Transformer (ViT) \cite{ircl21/vit} blocks and operates exclusively on the visible grids $\hat{\boldsymbol{f}}$. Specifically, the encoder first employs a trainable linear projection to patchify the input feature map to a sequence of $N$ vectors $\boldsymbol{z}_{1} \in \mathbb{R}^{N \times d_{1}}$, also referred to as grid tokens. To preserve the spatial information, position embeddings are added to the grid tokens. Then the visible tokens $\hat{\boldsymbol{z}}_{1}$ are encoded into the latent space by the ViT blocks to serve as the condition for the MAE decoder.

\paragraph{MAE Decoder.} The decoder processes the entire feature map $\boldsymbol{f}$ (or both the masked and visible grids), along with the condition obtained by the encoder $E_{M}(\hat{\boldsymbol{z}}_{1})$. The decoder also patchifies the input feature map to $N$ grid tokens $\boldsymbol{z}_{2} \in \mathbb{R}^{N \times d_{2}}$ and then adds the positional embeddings. Considering the diffusion process, the timestep $t$, which is closely related to the noise levels of the masked grids, can be optionally added to the grid tokens as well. Different from the ViT blocks used in \cite{cvpr22/mae}, we modify the decoder configuration to achieve better condition on the visible latents \cite{iccv23/diffmae}. We insert cross-attention layers before the self-attention layers in the ViT blocks for all the grid tokens, including both the masked and visible ones, to attend to the visible latents. Thus the grid tokens of different decoder blocks can attend to the corresponding visible latents of the different encoder blocks. Subsequently, all the gird tokens are processed by the remaining layers in the typical ViT blocks, namely multi-head self-attention and multi-layer perception. As depicted in Figure~\ref{fig:maediff_unet}, our MAE encoder and decoder adopt a U-shaped structure, which is known beneficial for image generation and dense prediction tasks.

The MAE mechanism functions as obtaining new grid tokens $\boldsymbol{z}$ from $D(\boldsymbol{z}_{2},E(\hat{\boldsymbol{z}}_{1}),t)$.
To make the MAE mechanism compatible with the diffusion U-Net, we first reshape the grid feature $\boldsymbol{z} \in \mathbb{R}^{N \times d_{2}}$ into the dimension of $d_{2} \times \frac{H}{r} \times \frac{W}{r}$, and then upsample it to $C \times \frac{H}{4} \times \frac{W}{4}$ using deconvolution layers. Consequently, the resulting feature map can be added to the U-Net encoder output feature $\boldsymbol{f} \in \mathbb{R}^{C \times \frac{H}{4} \times \frac{W}{4}}$, serving as input to the following bottleneck blocks and the U-Net decoder.

\section{Experiments}

\subsection{Datasets and Evaluation}

In our work, the publicly accessible dataset IXI \cite{ixi_dataset} is utilized to learn the healthy reference distribution during training. The IXI dataset, sourced from three different hospitals, comprises 560 pairs of brain MRI scans, each including T1-weighted and T2-weighted images. Our evaluation involves two other publicly available datasets: the 2021 Multimodal Brain Tumor Segmentation Challenge (BraTS21) dataset \cite{arxiv21/brast21_1,scidata17/brast21_2,tmi15/brast21_3} and the multiple sclerosis dataset from the University Hospital of Ljubljana (MSLUB) \cite{neuroi18/mslub}. The BraTS21 dataset contains 1251 MRI scans, including T1, post-contrast T1-weighted, T2-weighted, and T2-FLAIR modalities. The MSLUB dataset consists of 30 multiple sclerosis patients' MRI scans, including T1-weighted, contrast-enhanced T1-weighted, T2-weighted, and FLAIR modalities. Both the two datasets are supplemented with expert-annotated pixel-wise segmentation maps. We utilize the T2-weighted images of the three datasets for our experiments.

We employ the same data partition as in \cite{midl23/pddpm}. Specifically, for reconstruction testing, we utilize 158 T2-weighted images from the IXI dataset, while the remainder are allocated into five cross-validation folds. Each fold consists of 358 training images and 44 validation images, stratified according to the age of the patients. Regarding anomaly detection, the BraTS21 dataset, encompassing 1251 T2-weighted images, is divided into a validation set of 100 images and a test set of 1151 images. Similarly, the MSLUB dataset, encompassing 30 T2-weighted images, is split into a validation set of 10 images and a test set of 20 images.

We measure the reconstruction quality using the mean absolute reconstruction error ($l1$), and the anomaly detection performance using the average Dice score (DICE) and the average Area Under the Precision-Recall Curve (AUPRC).

\subsection{Data Preprocessing}

We apply a few preprocessing steps for all the T2-weighted images. First, each brain MRI image is aligned using affine transformations to match the SRI24-Atlas \cite{hum10/SRI24-Atlas}. We then proceed to skull stripping using HD-BET \cite{hum19/HD-BET}. It is noted that these procedures are already implemented in the BraTS21 dataset. After alignment and stripping, we trim the black borders around the scans, standardizing the resolution to $[192 \times 192 \times 160]$ voxels. A bias field correction is performed in the final step. For computational efficiency, we downsample the images by half, resulting in a reduced resolution of $[96 \times 96 \times 80]$ voxels, and eliminate 15 slices from both the top and bottom in the transverse plane.

\begin{table*}[!t]
    \centering
    \setlength{\tabcolsep}{2.5mm}{
    \begin{tabular}{lccccc} \toprule
    \multirow{2}{*}{Model}                       & \multicolumn{2}{c}{BraTS21}         & \multicolumn{2}{c}{MSLUB}     
                      & IXI     \\ \cmidrule(r){2-3} \cmidrule(r){4-5} \cmidrule(r){6-6}
                                                 & DICE (\%) $\uparrow$ & AUPRC (\%) $\uparrow$ & DICE (\%) $\uparrow$ & AUPRC (\%) $\uparrow$ & $l1 (1e-3)$ $\downarrow$ \\ \midrule
    Thresh \cite{miccai21/thresh}                & 19.69            & 20.27            & 6.21              & 4.23                        & 145.12         \\
    AE \cite{mia21/ae_vae}                       & 32.87$\pm$1.25   & 31.07$\pm$1.75   & 7.10$\pm$0.68     & 5.58$\pm$0.26     & 30.55$\pm$0.27 \\
    VAE \cite{mia21/ae_vae}                      & 31.11$\pm$1.50   & 28.80$\pm$1.92   & 6.89$\pm$0.09     & 5.00$\pm$0.40     & 31.28$\pm$0.71 \\
    SVAE \cite{midl22/svae}                      & 33.32$\pm$0.14   & 33.14$\pm$0.20   & 5.76$\pm$0.44     & 5.04$\pm$0.13     & 28.08$\pm$0.02 \\
    DAE \cite{midl22/dae}                        & 37.05$\pm$1.42   & \underline{44.99$\pm$1.72}   & 3.56$\pm$0.91     & 5.35$\pm$0.45     & \textbf{10.12}$\pm$\textbf{0.26} \\
    f-AnoGAN \cite{mia19/f-anogan}               & 24.16$\pm$2.94   & 22.05$\pm$3.05   & 4.18$\pm$1.18     & 4.01$\pm$0.90     & 45.30$\pm$2.98 \\
    AnoDDPM$^{\ast}$ \cite{cvprw22/anoddpm}      & 31.31$\pm$1.48 & 35.03$\pm$2.62 & 5.26$\pm$1.51 & 7.07$\pm$1.25 & 14.04$\pm$0.69 \\
    pDDPM$^{\ast}$ (s48) \cite{midl23/pddpm}     & 41.02$\pm$0.34 & 44.56$\pm$0.63 & \underline{9.57$\pm$1.03} & \textbf{9.12}$\pm$\textbf{0.64} & \underline{11.35$\pm$0.16} \\ 
    pDDPM$^{\ast}$ (s16) \cite{midl23/pddpm}     & \underline{41.10$\pm$0.75} & 44.06$\pm$0.63 & 8.78$\pm$0.92 & \underline{9.04$\pm$0.51} & 12.03$\pm$0.43 \\ 
    MAEDiff (Ours)                               & \textbf{42.07}$\pm$\textbf{1.12}   & \textbf{45.48}$\pm$\textbf{1.47}   & \textbf{9.61}$\pm$\textbf{0.83}     & \underline{9.04$\pm$0.58}     & 11.48$\pm$0.56 \\ \bottomrule
    \end{tabular}}
    \caption{Quantitative comparison of the state-of-the-art approaches with our MAEDiff. We conduct our experiments over five folds and present the results as the mean $\pm$ standard deviation for all metrics. $^{\ast}$ denotes our reimplementation. The \textbf{bolded} and \underline{underlined} results denote the best and second-best results, respectively.}
    \label{tab:model_comparison}
\end{table*}

\begin{table*}[!t]
    \centering
    \setlength{\tabcolsep}{4.5mm}{
    \begin{tabular}{lccccc} \toprule
    \multirow{2}{*}{Model}                       & \multicolumn{2}{c}{BraTS21}         & \multicolumn{2}{c}{MSLUB}     
                      & IXI     \\ \cmidrule(r){2-3} \cmidrule(r){4-5} \cmidrule(r){6-6}
                                                 & DICE (\%) $\uparrow$ & AUPRC (\%) $\uparrow$ & DICE (\%) $\uparrow$ & AUPRC (\%) $\uparrow$ & $l1 (1e-3)$ $\downarrow$ \\ \midrule
    U-Net                                        & 41.10$\pm$0.75   & 44.06$\pm$0.63   & 8.78$\pm$0.92     & 9.04$\pm$0.51     & 12.03$\pm$0.43 \\
    U-Net $+$ Att.                               & 15.87$\pm$0.02   & 9.71$\pm$0.38    & 4.11$\pm$1.17     & 2.64$\pm$0.07     & 205.39$\pm$31.89$\times 10^{3}$ \\
    U-Net $+$ MAE                                & 41.29$\pm$1.76   & 44.67$\pm$2.18   & 9.60$\pm$1.28     & 8.94$\pm$0.71     & 11.41$\pm$0.33 \\ 
    U-Net $+$ Att. $+$ MAE                       & 42.07$\pm$1.12   & 45.48$\pm$1.47   & 9.61$\pm$0.83     & 9.04$\pm$0.58     & 11.48$\pm$0.56 \\ \bottomrule
    \end{tabular}}
    \caption{Ablation study of different modules. Att. denotes the global attention in the diffusion U-Net encoder and decoder. MAE denotes our proposed MAE mechanism.}
    \label{tab:ablation_component}
\end{table*}


\subsection{Implementation Details}

We implement all the experiments based on \cite{midl23/pddpm}. Following \cite{cvprw22/anoddpm,midl23/pddpm}, we employ the Simplex noise \cite{tg02/simplex} instead of the Gaussian noise in the diffusion process due to its superior capability to corrupt the anomalous components in brain images. The variance of noise follows a linear schedule ranging from $1e-4$ to $2e-2$. During the training phase, the timesteps are uniformly sampled from the interval $[1,1000]$. For the testing phase, we fix the timestep at $t_{test}=500$.

In our implementation, we set the patch size to $p=48$, patch stride to $s=16$, and grid size to $r=16$. We use the same U-Net architecture as in \cite{midl23/pddpm}, which is modified from the U-Net in \cite{NeurIPS21/guided_diffusion}. 
Additionally, timestep embeddings are projected into appropriate dimensions and integrated into each block of the U-Net. 

With regards to the MAE mechanism, we use the standard ViT \cite{ircl21/vit} blocks for the MAE encoder. 
Specifically, we employ the small ViT model as our MAE encoder, which has 12 ViT blocks with hidden size $d_{1}$ of 384 and 6 attention heads. 
For the MAE decoder, we add a cross-attention layer to every ViT block. The MAE decoder has 8 blocks with hidden size $d_{2}$ of 512 and 16 attention heads. These MAE decoder blocks uniformly attend to the outputs of the MAE encoder blocks. We use two distinct linear projections for the encoder and decoder to obtain the grid tokens. The fixed sinusoidal position embeddings are added to the tokens. We do not integrate the timestep embeddings in our MAE mechanism.

We train all the models using the Adam optimizer for up to a total of 1600 epochs. The best checkpoint for each model, which is identified based on the performance on the validation dataset, is utilized for the testing phase. The training batch size is 32 and the learning rate is set to 0.0001. All experiments are conducted using the PyTorch framework on a single NVIDIA GeForce RTX 3090 GPU with 24GB memory.
\begin{figure*}[!t]
    \includegraphics[width=7in]{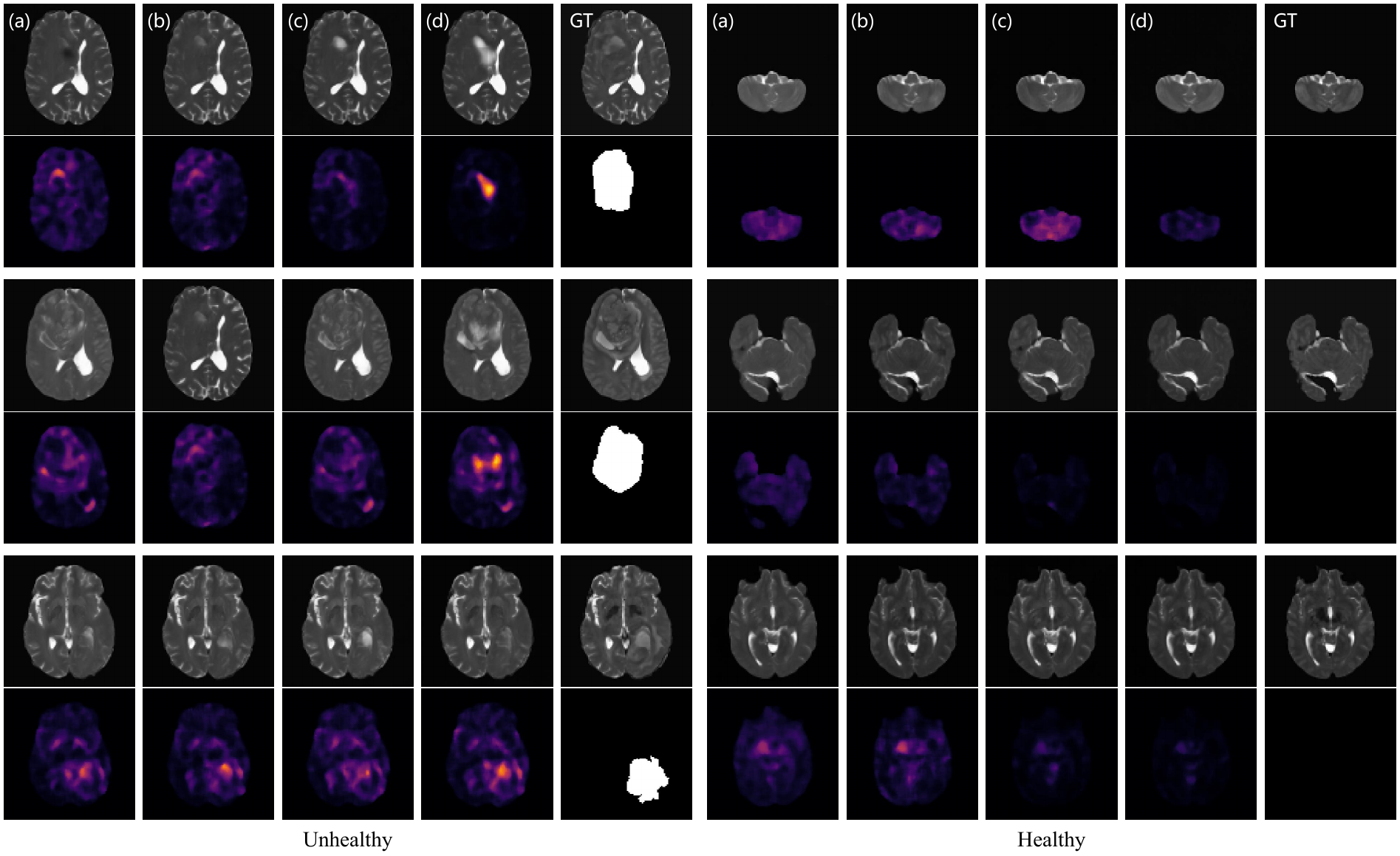}
    \caption{Qualitative comparison of our MAEDiff and previous approaches. Columns (a)-(d) display the reconstruction results and the anomaly score maps for the AnoDDPM, pDDPM (s48), pDDPM (s16), and our MAEDiff, respectively. Column GT presents the original images and their corresponding ground truth annotations for reference.}
    \label{fig:vis}
\end{figure*}

\subsection{Post-Processing and Anomaly Scoring}

Recall our underlying assumption: by noising the diseased images, the pathological areas can be corrupted, and these images can be reconstructed to their healthy counterfactuals without significant modifications by the model pre-trained merely on the healthy images. Thus by comparing the generated healthy counterfactuals with the original diseased ones, we can identify the anomalies in areas where the reconstruction errors are high. The $l1$ reconstruction error can be regarded as a pixel-wise anomaly score $\Delta_{AS} = \vert \boldsymbol{x}_{0} - \boldsymbol{x}_{0}^{rec} \vert$. Following \cite{midl23/pddpm}, our methodology includes several post-processing steps, as typically found in existing literature \cite{mia21/ae_vae,midl19_/context_vae}. This involves smoothing $\Delta_{AS}$ with a median filter of kernel size $K_{M} = 5$ and applying brain mask erosion for three cycles. Post-binarization of $\Delta_{AS}$, we engage in a connected component analysis to exclude segments smaller than 7 voxels. Thereinto, the threshold for binarization of $\Delta_{AS}$ is ascertained through a greedy search on the validation dataset comprised of unhealthy samples, optimizing the Dice score across various thresholds. The optimal threshold identified is then utilized to compute the average Dice score (DICE) on the unhealthy test set. Additionally, we calculate and present the average Area Under the Precision-Recall Curve (AUPRC) and the mean absolute reconstruction error ($l1$) on the healthy test split of our IXI dataset.

\subsection{Results and Analyses}


\paragraph{Comparison with Existing Approaches.} We compare the performance of our MAEDiff against several existing approaches, including Thresh \cite{miccai21/thresh}, AE, VAE \cite{mia21/ae_vae}, SVAE \cite{midl22/svae}, DAE \cite{midl22/dae}, f-AnoGAN \cite{mia19/f-anogan}, AnoDDPM \cite{cvprw22/anoddpm} and pDDPM \cite{midl23/pddpm}. Notably, we reimplement the AnoDDPM and pDDPM using the same U-Net architecture and training scheme as used in our MAEDiff for a fair comparison. The pDDPM uses the same patch size of $p=48$ as our MAEDiff, and two distinct patch strides $s=48$ and $s=16$. As demonstrated in Table~\ref{tab:model_comparison}, our proposed MAEDiff model significantly surpasses all previous approaches, achieving nearly a 1$\%$ improvement on the BraTS21 dataset in terms of both DICE and AUPRC metrics. On the MSLUB dataset, MAEDiff continues to demonstrate a moderate yet notable superiority in performance. In terms of the reconstruction quality, evaluated by the $l1$ error on the IXI dataset, the VAE performs the best and marginally outperforms ours. However, the anomaly detection performance of the VAE is not satisfactory.

\paragraph{Visualization.} We visualize the reconstruction results and the anomaly score maps for our MAEDiff and other approaches on dataset BraTS21 in Figure~\ref{fig:vis}. It can be seen that on unhealthy brain images, our MAEDiff demonstrates a notable capacity to accurately reconstruct brain anatomical structures, even when significant portions of critical brain structures are distorted. This is evident from the first two examples, where other approaches either fail to correctly reconstruct these structures or erroneously retain the diseased elements. The third example further highlights our method's effectiveness in eliminating pathological content, showcasing its superior performance in handling complex anatomical variations. In terms of reconstruction quality on healthy images, our MAEDiff distinctly excels in preserving intricate details and original contrasts, leading to superior results.


\paragraph{Effectiveness of Different Modules.} In Table~\ref{tab:ablation_component}, we perform several architectural ablation studies to assess the efficacy of our proposed MAE mechanism. We implement the U-Net baseline, U-Net using global attention at $48 \times 48$ and $24 \times 24$ resolutions, U-Net using MAE mechanism, and U-Net using both global attention and MAE mechanism. The U-Net baseline is equivalent to the pDDPM (s16) in the main table, which does not outperform the pDDPM (s48). This implies its deficiency in integrating global information, leading to inconsistent reconstructions of overlapping patches and potential performance degradation when these patches are averaged. Interestingly, merely adding global attention to the diffusion U-Net surprisingly leads to a decrease in the model performance. This could be attributed to the adverse impacts of applying global attention operations across a combination of noised and clean regions. In contrast, our proposed MAE module, as indicated by all metrics, significantly enhances performance compared to the baseline. This improvement underscores the MAE module’s effectiveness in assimilating global information, thereby ensuring consistent reconstructions of overlapping patches, and better conditioning on the visible regions. These improvements are crucial in boosting both anomaly detection accuracy and reconstruction quality. Moreover, incorporating global attention in the diffusion U-Net encoder and decoder further amplifies this performance enhancement.


\section{Conclusion}

In this work, we present a Masked Autoencoder-enhanced Diffusion Model (MAEDiff), aiming at reconstructing the healthy references by patch-based denoising. MAEDiff utilizes a hierarchical patch partition strategy, where the upper-level overlapping patches are sequentially denoised to more accurately reconstruct the fine details, and the sub-level non-overlapping patches are processed using an MAE-like mechanism to enhance the conditional generation. Experimental results attest to the effectiveness of our MAEDiff. In the future, we intend to integrate the selection of noised patches with a preliminary lesion identification technique, which could further leverage the strengths of masked modeling. Additionally, we plan to explore cross-domain methods to enhance the model’s generalizability across different test datasets.


\bibliographystyle{named}
\bibliography{ijcai24}

\end{document}